\documentclass[journal, a4paper]{IEEEtran}

\usepackage{hyperref}
\usepackage{amsfonts}
\usepackage{cite}      

\usepackage{graphicx}   

\usepackage{url}        

\usepackage{amsmath}    

\usepackage{siunitx}
\newcommand{\Rev}[1]{\textcolor{black}{#1}}

\begin{document}

	\title{From reductionism to realism: Holistic mathematical modelling for complex biological systems}
	\author{Ramón Nartallo-Kaluarachchi$^{1,2,3}$, Renaud Lambiotte$^{1}$, and Alain Goriely$^{1}$ \\ \small $^{1}$Mathematical Institute, University of Oxford, Woodstock Road, Oxford, OX2 6GG, United Kingdom\\
$^{2}$Centre for Eudaimonia and Human Flourishing, University of Oxford, 7 Stoke Pl, Oxford, OX3 9BX, United Kingdom\\
\textit{Correspondence to} \texttt{\{nartallokalu\}\{lambiotte\}\{goriely\}@maths.ox.ac.uk}}
	\maketitle

\begin{abstract}
At its core, the physics paradigm adopts a reductionist approach, aiming to understand fundamental phenomena by decomposing them into simpler, elementary processes. While this strategy has been tremendously successful in physics, it has often fallen short in addressing fundamental questions in the biological sciences. This arises from the inherent \textit{complexity} of biological systems, characterised by heterogeneity, polyfunctionality and interactions across spatiotemporal scales. Nevertheless, the traditional framework of complex systems modelling falls short, as its emphasis on broad theoretical principles has often failed to produce predictive, empirically-grounded insights. To advance towards actionable mathematical models in biology, we argue, using neuroscience as a case study, that it is necessary to move beyond reductionist approaches and instead embrace the complexity of biological systems—leveraging the growing availability of high-resolution data and advances in high-performance computing. We advocate for a holistic mathematical modelling paradigm that harnesses rich representational structures such as annotated and multilayer networks, employs agent-based models and simulation-based approaches, and focuses on the inverse problem of inferring system dynamics from observations. We emphasise that this approach is fully compatible with the search for fundamental biophysical principles, and highlight the potential it holds to drive progress in mathematical biology over the next two decades.
\end{abstract}

\section{Introduction}
A fundamental aspect of human nature and success as a species is the desire to make sense of the world around us. Over the course of thousands of years and through the scientific revolution, mathematics, specifically \textit{mathematical modelling} (MM), has become the dominant and ultimate paradigm used to abstract physical systems and describe their behaviour over time and space. This has resulted in the unparalleled success of MM in domains like physics and engineering, and has profoundly changed human society. The relationship has become naturally symbiotic with new mathematics constantly being discovered through the development of models of real-world phenomena. Many such examples have become canon in the field of MM such as Poincaré's attempts to model celestial mechanics \cite{Poincaré1892celestialmech}, which laid the ground-work for the study of dynamical systems, divergent series, and resonance phenomena. In an effort to understand the bridges of Koenigsberg, Euler developed concepts that would later culminate in graph and network theory \cite{Euler1741bridges}. More recently, whilst deriving simple equations to describe the dynamics of the climate, Lorenz discovered the first example of a chaotic attractor \cite{Lorenz1963chaos}. Whilst we are still far from a complete theory, it can be argued that 20$^{\text{th}}$ century was the century of physics when mathematical abstractions  led to a significant maturity in the field, with many fundamental problems being solved and theories confirmed \cite{Agar2012science}.
\begin{figure*}
   \centering
    \includegraphics[width=0.8\linewidth]{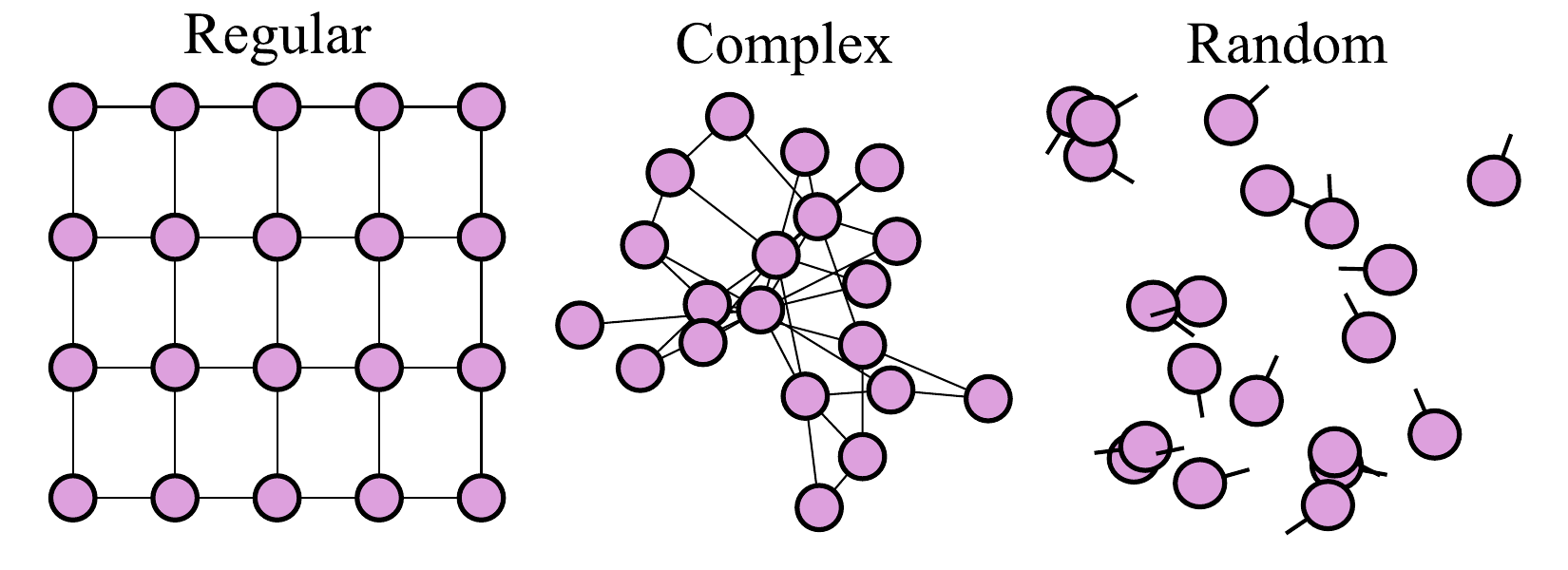}
    \caption{\textbf{The spectrum of interaction structures in mathematical models.} Models of physical system exist on a spectrum from the regular to the random. For regular structures, continuum assumptions are often valid leading to a low number of degrees of freedom. In the random case, behaviour is typically described by a statistical distribution rather than individual elements.  Between these extrema are complex systems, where interactions are heterogeneous leading to high-dimensional dynamics.}
    \label{fig: spectrum} 
\end{figure*}
However, biological systems, spanning from entire ecosystems to sub-cellular processes, have proved to be more impregnable to mathematical analysis. This is why it was first developed as a mostly descriptive science before moving to theoretical concerns. Yet, many fundamental theoretical questions in biology remain unanswered. How does life emerge from a precise arrangement of individually inanimate atoms? How does evolution shape species? How are cognition and consciousness orchestrated by the noisy, electrical signals of the human brain? How can we leverage the mechanisms of biology to eradicate disease? These problems demand solutions and, as of yet, mathematics has failed to provide a clear and consensus framework promising to yield them.\\
\\
Nevertheless, it would be inaccurate to characterise mathematical biology as an immature discipline, as numerous models have demonstrated great success and provided valuable insights. However, these breakthroughs often exist in isolation and are typically limited to specific biological systems or processes. To understand why this limitation arises from the intrinsic nature of biological systems, we must first identify the characteristics of the dominant \textit{reductionist} approach that has resulted in such rampant success in physics. Throughout this \textit{Perspective}, we will build our arguments with examples primarily from neuroscience and the human brain,  which serves as the prototypical example of a complex biological system due to its inherent complexity, intricate organisation, and central importance to all aspects of human life and society.\\
\\
We start with one of the most important breakthroughs in mathematical biology, the Hodgkin-Huxley (HH) model for electrophysiological excitability \cite{Hodgkin1952membrane}. Some species of squid possess a so-called `giant axon', up to 1.5mm in diameter, that controls water propulsion \cite{Young1938squid}. By isolating this large axon, Hodgkin and Huxley in the 1950s were able to insert voltage clamp electrodes and vary extracellular ionic concentrations to develop a complete set of \textit{ordinary differential equations} (ODEs) that describe the propagation of electrical signals through a neuronal axon, providing the first complete description of the action potential, the basis of all neuronal communication. This is a canonical example of \textit{reductionism}, a philosophical approach that assumes that complex phenomena can be described in terms of a series of simpler, more fundamental processes. In the case of the HH model, the key assumption was that neuronal activity could be modelled as an electrical circuit, with each component of the cell being represented by an electrical element. This, in turn, abstracts away other subcellular and biochemical processes.\\
\\
The reductionist approach has proved extremely successful across science, but particularly in physics and chemistry, where simple experiments with only a few independent variables can reveal fundamental principles, and fundamental principles can be assembled together like building blocks to explain more complex situations. However, this approach may also be responsible for the lack of theoretical success that has plagued the biological sciences \cite{Nagel2012mindandcosmos,Anderson1972MoreIsDifferent,Ball2025complexity,Brenner2010sequences,Gyllingberg2023lostart}. For example, whilst there are only a few elementary units for physical systems e.g. protons, neutrons and electrons, biological systems are characterised by an array of unique elements e.g. the zoo of different lymphocytes in the immune system, the collections of proteins, or the extremely diverse populations of neurons and glial cells in the nervous system. Returning to Hodgkin and Huxley, whilst their result represented the crucial breakthrough in understanding the subcellular mechanisms governing neural firing, it brought us only marginally closer to understanding cognition or computational processes in the brain.\\\\
It is well known that many physical systems can be organised onto a spectrum spanning from the \textit{regular} to the \textit{random} \cite{watts1998smallworld}, as illustrated in Fig. \ref{fig: spectrum}. It is well appreciated that the reductionist approach appears to be most effective at the extrema of this spectrum, where interactions present certain types of symmetries. At one end of the spectrum, for systems characterised by regular, symmetric organisation, such as atoms in a metal or crystal, individual elements can be abstracted away through continuum assumptions. For example, the flow of heat can be modelled using a \textit{partial differential equation} (PDE) framework, without a need to model individual atoms \cite{Fourier1822heat}. At the other end of the spectrum, in order to describe random systems composed of enormous numbers of noisy, interacting components, such as the behaviour of molecules in a gas, physicists developed the framework of \textit{statistical mechanics} \cite{Gibbs1902statmech}. Under this framework, it is typically assumed that interacting elements are identical. Moreover, one does not describe the behaviour of each individual element but instead the dynamics of a probability distribution that describes the ensemble, typically through mean-field approximations. This reduces the number of degrees of freedom and renders tractable mathematical descriptions. \\
\\
Between these two extremes lie systems that, for lack of a better definition, are best described as \textit{complex}. Such systems are characterised by topologically irregular interactions, heterogeneous elements, poly-functional processes, and evolve over multiple time- and length-scales. These features are particularly ubiquitous in biological processes. Returning to the brain, consider the activity in a large population of interconnected neurons. An accurate model would incorporate the complex connectivity between units, the heterogeneity of different cell-types, as well as the neurochemical processes underpinning their mechanism. Consequently, modelling their behaviour would require an intricate model with an enormous number of degrees of freedom, invariably leading to high-dimensional dynamics. Of course, the idea that biology is characterised by its complexity is not new \cite{Schrodinger1944whatislife,GellMan1995quark}, and much of the emerging field of \textit{complex systems science} is motivated by the study of biological systems \cite{Maayan2017complex,Krakauer2024complexworld}. Nevertheless, despite the sustained attention over the past two decades, complex systems approaches to biology have yet to yield convincing and tangible results.\\
\\
In this \textit{Perspective}, we focus on the problem of biological complexity and MM. \Rev{In particular, we present a unified framework of complementary research directions, which together form what we call \textit{holistic mathematical modelling}. We make the case that in modern sciences the classical \textit{reductionist approach} cannot work in isolation, and it is necessary to develop models that consolidate different spatial and temporal scales, use multimodal data, and consider complex representational structures. This holistic approach goes} beyond isolated functions to integrate multiple biophysical processes within a single model. Moreover, we argue that the theoretical approaches to complex systems that have been dominant for the past two decades\Rev{, especially those models that are inspired by biological problems, are} limited and often cannot yield empirically useful or actionable insights. \Rev{Yet, the extent of the data and computational requirements needed for high-fidelity models is often prohibitive. Despite this, we believe that the development of such models is a necessity for progress in mathematical biology to match the accomplishments of experimental biology.} Looking to the future, we argue that modern tools in data collection, \textit{machine learning} (ML), \textit{artificial intelligence} (AI), and \textit{high-performance computing} (HPC) will not replace modelling, but instead facilitate it at unprecedented scale. By adopting a modern approach, we present a holistic modelling framework that embraces complexity and empiricism\Rev{, whilst synergising with exisiting approaches, to develop a collective arsenal capable of answering} the most pressing scientific challenges.

\section{Rich and realistic representations for complex systems}
\begin{figure*}
    \centering
    \includegraphics[width=\linewidth]{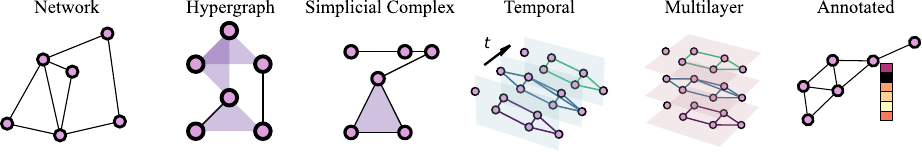}
    \caption{\textbf{Richer representations for complex systems.} Networks have become the dominant paradigm for representing interactions in complex systems. However, a range of extensions exist that have the expressiveness to represent more complex interactions as well as their dynamics. Examples include hypergraphs, simplicial complexes, as well as temporal, multilayer and annotated networks.}
    \label{fig: rich}
\end{figure*}
In an effort to accurately represent real-world systems, from biology and beyond, the study of \textit{networks} has become one of the most popular methods for abstracting elements and their interactions \cite{newman2018networks}. In the last 20 years, this has lead to a number of important studies such as that of `scale-free' \cite{barabasi1999pa}, `small-world' \cite{watts1998smallworld}, or community-structured networks \cite{Girvan2002community}. As a result, the study of complex systems has become, for many, synonymous with \textit{network science}. This abstraction has been particularly useful in the modelling of a range of biological systems such as the brain \cite{basset2017networkneuro}, ecosystems \cite{dunne2002ecology}, biomedical systems \cite{barabasi2011networkmedicine}, protein interactions \cite{Safari2014protein} and gene regulation \cite{Davidson2005gene}. By introducing weighted, directed and signed connections, as well as \textit{temporal networks}, where links vary in time, these abstractions can be extended to model a variety of different biological processes.\\
\\
Whilst intrinsically motivated by applications, the study of complex systems has, to date, remained largely theoretical, with a substantial focus on general principles that aim to transcend the confines of a specific system or process \cite{Bianconi2023complexsystems}. \Rev{Similarly, many studies aimed directly at biological complexity, focus on `top-down' models, which aim at reproducing qualitative \textit{emergent} phenomena \cite{Pezzulo2016topdown}, as opposed to predictive accuracy. Whilst this is no doubt important, it has broadly failed to represent biological systems with the required authenticity to posit or confirm theories, or often to yield practical results. As a result, complex systems science can often exist in a vacuum, adjacent to the mainstream of biology.\\
\\
In recent times, in an effort to generalise networks to more expressive, realistic structures,} significant attention has turned towards \textit{higher-order} representations of complex systems \cite{lambiotte2019higherorder,Torres2021whyhow,battinson2020physics} such as \textit{hypergraphs}, \textit{simplicial complexes} and \textit{multilayer networks}, illustrated in Fig. \ref{fig: rich}. \Rev{Multilayer networks have proved to be particularly useful representations for multi-scale biological systems \cite{Gosak2018network} as they allow for modelling a number of different, yet coupled processes, evolving simultaneously \cite{Kivelä2014multilayer,Betzel2017multiscale}}. For example, neural activity in the human brain network is facilitated by blood-flow through the cerebral vasculature. Moreover, ageing results in structural changes to the networks that facilitate cognition. Whilst these processes have different mechanisms, dynamics and time-scales, they remain intrinsically coupled, reducing the efficacy of any model that attempts to consider them in isolation. \Rev{Despite this, many models on multilayer networks remain largely theoretical, as they are typically overparameterised, requiring large amount of multimodal data for fitting. Nevertheless, the increasing availability and quality of multimodal data raises the opportunity for this expressive modelling technique to be integrated into more powerful, predictive methods, as shown by the recent \textit{MINT} toolbox for neuroimaging data in disease \cite{Sarraf2025MINT}. We believe that the next 20 years will see an increased focus on this type of data and modelling.}\\
\\
Somewhat surprisingly, one higher-order representation of a complex system whose dynamics are rarely considered is the \textit{annotated network}\footnote{It is worth noting that whilst annotated networks have been rarely studied in terms of dynamics, they are prevalent in some fields, such as \textit{graph-representation learning} \cite{Hamilton2020GRL}.} \cite{Newman2016annotated}. Unlike the simple structure found in graph theory of a tyical network, $G = (V,E)$, defined by a vertex set, $V$, and edge set, $E$, annotated networks can encode nodal \textit{metadata} in the form of an additional vector of features $\mathbf{d}_i \in \mathbb{R}^d$ that is attached to each vertex $i\in V$ (or to each edge) \cite{Peel2017metadata,Bassolas2022metadata}. In particular, this extension allows us to model the heterogeneity of individual elements, adding a crucial layer of additional expressiveness to our mathematical models. Whilst still nascent, such an approach has proved useful in neuroscience in the form of \textit{biologically annotated connectomes} \cite{Bazinet2023biologicallyannotated}, which are human brain networks obtained from structural imaging where each node (region) is imbued with additional information such as cell-type density, gene transcription, receptor density, and myelination. Open-source software, such as \textit{neuromaps} \cite{Markello2022neuromaps}, facilitates the integration of this \textit{multimodal} \textit{data} into richer structural representations. Such heterogeneity has been used to develop biophysical models of whole-brain dynamics that incorporate regional heterogeneity through variations in gene expression \cite{Deco2021dynamical} or synaptic strength \cite{Demirtaş2019heterogeneity} for added biological complexity and closer agreement with observed neural dynamics.\\
\\
\Rev{In an attempt to tackle the complexity of biological systems, the time has come to apply these frameworks more concretely to problems in biology. We are not arguing here that complex systems approaches to biology are entirely novel, but instead that the primary focus on emergent phenomena and the emphasis on universal phenomena have limited the accuracy and predictive power of these models. In turn, the mainstream within these fields may replace reductionist abstractions with richer representations that are capable of incorporating the heterogeneity seen in experiments. To achieve this goal, integration with multimodal data is crucial, but it is worth noting the scaling challenge that has proved limiting to this point. Given a system with $N$ elements, the number of pairwise interactions scales as $N^2$. If we introduce higher-order or multilayer interactions, the parameter space scales at an even faster rate  $\sim N^b$. As a result, such over-parametrised models are plagued by both a lack of sufficient multimodal data and the inference algorithms needed to fit them.} However, as we progress into the age of \textit{big data}, modern experimental techniques are able to record multimodal data on an unprecedented scale, particularly in biological systems \cite{Marx2013bigdata,Sejnowski2013neurobigdata}. Moreover, novel techniques in computational statistics and ML are able to perform parameter inference for such complex models using observed data \cite{Gaskin2023neural,Cranmer2020simulation}. \Rev{Unlike the purely statistical models typically found in ML, these systems have parameters that remain both physically constrained and meaningful. This ensures the interpretability of complex dynamical models and allays the usual concern about von Neumann's elephant\footnote{Whilst often attributed to von Neumann, the quote ``...with four parameters I can fit an elephant, and with five I can make him wiggle his trunk.'' first appears in an article by Freeman Dyson, where he quotes Enrico Fermi, who is in turn quoting von Neumann \cite{Dyson2004meeting}.} and his wiggly trunk \cite{Dyson2004meeting}.} With accurate calibration on experimental data and efficient simulation on HPCs, detailed, heterogeneous and elaborate models of complex systems can make the leap from generic and descriptive to mechanistic and predictive.
\subsection{Integrating many time- and length-scales}
\begin{figure*}
    \centering
    \includegraphics[width=0.6\linewidth]{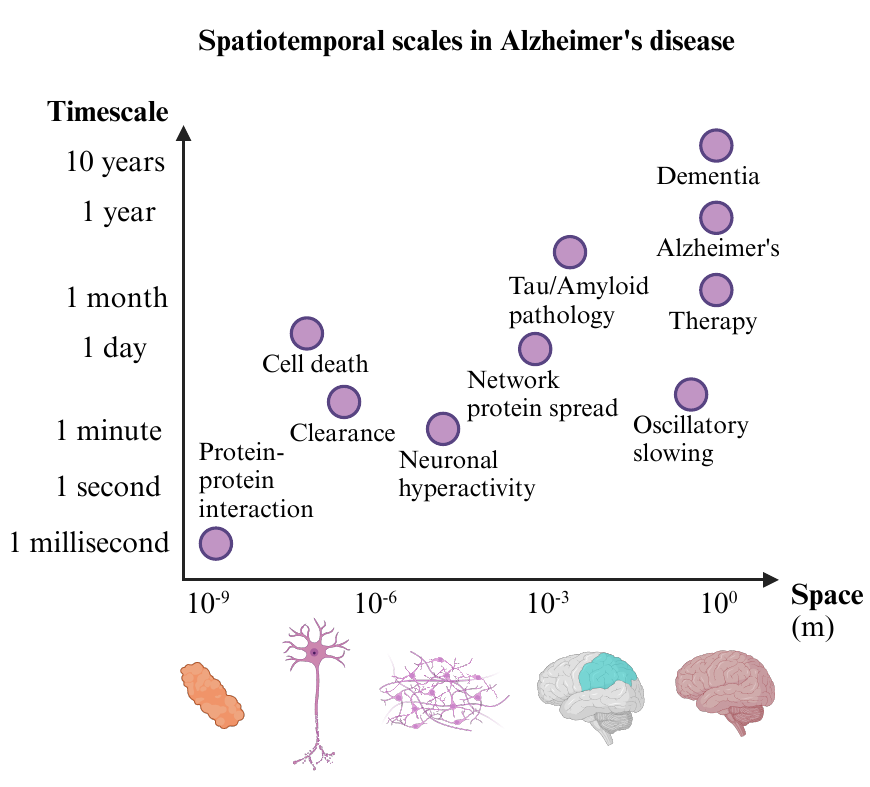}
    \caption{\textbf{Spatiotemporal scales in Alzheimer's disease.} Complex biological processes, such as the development of neurodegenerative diseases such as Alzheimer's disease (AD), evolve over a range of interacting spatiotemporal scales. Focussing on AD, protein misfolding ultimately leads to the spread of toxic tau and amyloid-$\beta$ proteins through axonal fibres, which is partially cleared through cell regulation. Ultimately, this leads to cell death and pathological neural dynamics. This manifests in tau and amyloid pathology and, finally, cognitive decline which can be combatted with therapy. A MM of this process must take into account the full range of spatiotemporal scales in order to build a mechanistic understanding of AD.}
    \label{fig: AD}
\end{figure*}

Whilst rich structures such as multilayer, annotated, or hyper-networks offer the expressiveness necessary to model a range of interacting biological processes, one remaining challenge is that many of these processes evolve over multiple time- and length-scales. In domains like physics, such scales of interaction can often be separated, hence the success of reductionist single-scale models. For example, to model the motion of planets in a solar system, we do not need to consider the atmospheric dynamics that occur at a smaller spatial scale or the gravitational field of stars that are light-years away. Similarly, a celestial model can assume that stars and planets are of fixed mass, ignoring stellar evolution and possible collision events, which occur on a much slower time-scale, thus are insignificant. These processes are mostly decoupled which allows for a remarkably precise modelling of planet motions over millennia. \\
\\
The situation is completely different for biological systems that are best characterised by interacting spatiotemporal scales \cite{Dada2011multiscale}. Focusing on a single scale, naturally limits our ability to explain certain phenomenon. For example, neurodegenerative diseases, like Alzheimer's disease (AD), evolve due to the spread of toxic proteins over the course of many years \cite{Soto2003unfolding}. However, the symptoms of these conditions manifest themselves in cognitive decline stemming from pathological neural activity, on the order of $10^{-3}-10^0$ seconds \cite{Haan2012activity}. In order to understand the mechanisms by which AD causes cognitive decline, it is insufficient to only model either neural activity or toxic protein spread in isolation. Fig. \ref{fig: AD} highlights the many processes underlying the development of AD along with their respective time- and length-scales. Recent studies have presented an approach that attempts to integrate these time-scales \cite{Goriely2020neuronal,Alexandersen2023multiscale}. First, they model network degradation due to toxic-protein spread, freezing the state of the network at distinct snapshots to then form the topology underlying a model of fast neural activity.\\
\\
Whilst rich representations of complex systems can capture the intricate structure of interactions in biological processes, the full characterisation of their dynamics requires multi-scale models that couple interacting processes. As with many of these proposed frameworks, this naturally complicates mathematical analysis and elegance in favour of greater biological realism.
\section{Towards simulation-based science}
\subsection{Agent-based models}
Mathematical models are often appreciated for their elegance and parsimony. In particular, such models can often be treated analytically leading to great insight. As models become more complex, they lose analytical tractability and numerical methods of analysis must be used.
One framework for modelling complex systems that hinges on direct simulation, is the use of \textit{agent-based models} (ABMs), which have been embraced in economics \cite{Axtell2022abm,Farmer2009abm},  computational social science \cite{Volkening2022primer}, \Rev{and mathematical biology} \cite{An2009agentbasedmodel,goriely17}.\\
\\
The typical structure of ABMs is the time evolution of a micro-scale model that defines the characteristics of individual elements and their interactions. Examples of such systems include models of the spread of disease through populations  and ecological dynamics or collective motion with agents being individuals or populations (Panel A, Fig. \ref{fig: sim}). Unlike traditional models, ABMs call for a different approach to mathematical analysis, one that is often realised through exhaustive simulation and data-driven calibration. For this reason, such models were previously found to be computationally restrictive. However, following advancements in scientific computing, large-scale simulations of ABMs have become both tractable and attractive \cite{Bodine2020agentbased}.\\
\\
The adoption of ABMs requires a shift in attitudes towards mathematical models. In particular, the proponents of such models argue that the complexity of many systems requires a shift towards a purely \textit{simulation-based science}, where  additional features and realism are favoured over parsimony and elegance \cite{Lavin2022simulation}. The current inclination toward a minimal model is often justified via \textit{Occam's razor}, the philosophical argument that the simplest model is often the best. However, the lack of success of such minimal models in biology, compounded with the success of overparameterised ML models, has called into question this long-standing perspective \cite{Ball2016occam,Dubova2025ockham}. Whilst ML models are able to discover a multiplicity of latent features that have predictive power, a parsimonious mathematical modeller must instead identify  the handful that may be important. Whilst not without critics \cite{Mazin2022inverseoccam}, more complex models relieve the modeller of some of this choice by allowing them to include sufficient complexity at the level of model definition. Moreover, it is important to note the differences between an ABM and a ML model. Whilst both lack analytical tractability and rely on computationally-intensive processes, the latter may be considered to be a `black-box', whose mechanisms remain somewhat opaque, whilst the former attempts to model the basic physical mechanisms governing a system. This poses additional advantages, such as the ability to test realistic interventions in a system due to the model's interpretability \cite{Baker2018mechanistic}.\\
\\
Let us consider the example of an epidemic model. A traditional compartmental model, such as the `SIR' model \cite{Kermack1927SIR}, is clearly overly simplistic and cannot express the heterogeneity of individuals in a population. Yet, it is mathematically tractable due to its ODE formulation and can be analysed to obtain fundamental insight and identify key quantities such as the \textit{basic reproduction number}, $R_0$. Moreover, with simple parameter modifications, the model can incorporate the effect of interventions such as social distancing or medication. On the other hand, \textit{deep learning} based models are able to analyse large amounts of data, such as digital, genomic, environmental and behavioural data, in order to make accurate predictions about the development of the epidemic \cite{Rodríguez2024machineepidemic}. However, we cannot test the effects of interventions in such a model due to its opacity. Finally, we can also consider an ABM, which can model population heterogeneity, experiment with a variety of interventions, and make realistic predictions. Again, one major drawback is the need for exhaustive simulation and data for calibration. With modern digital data collection methods and improved HPC, though, such restrictions will loosen in the coming years.\\
\\
Despite these advantages, the necessity to specify many local rules of interaction can cause ABMs to suffer from \textit{ad-hoc} selection of many \textit{hidden} parameters. Typically at the modeller's discretion, such parameters are chosen inline with expected behaviour, leading to biases towards modeller's hypotheses. The modeller may often tinker with local rules until an expected behaviour is found on the screen, leading to the danger that the modelling process resembles a video game with the sole purpose of mimicking reality rather than understanding emergent behaviours.  In search of a transparent framework for ABMs, modellers should aim to minimise hidden parameters and choose priors in a principled, rational manner rather than performing endless hand-tuning. These, in turn, may be updated with a Bayesian mechanism, leveraging available data. 
\subsection{Digital twins}
\begin{figure*}
    \centering
    \includegraphics[width=\linewidth]{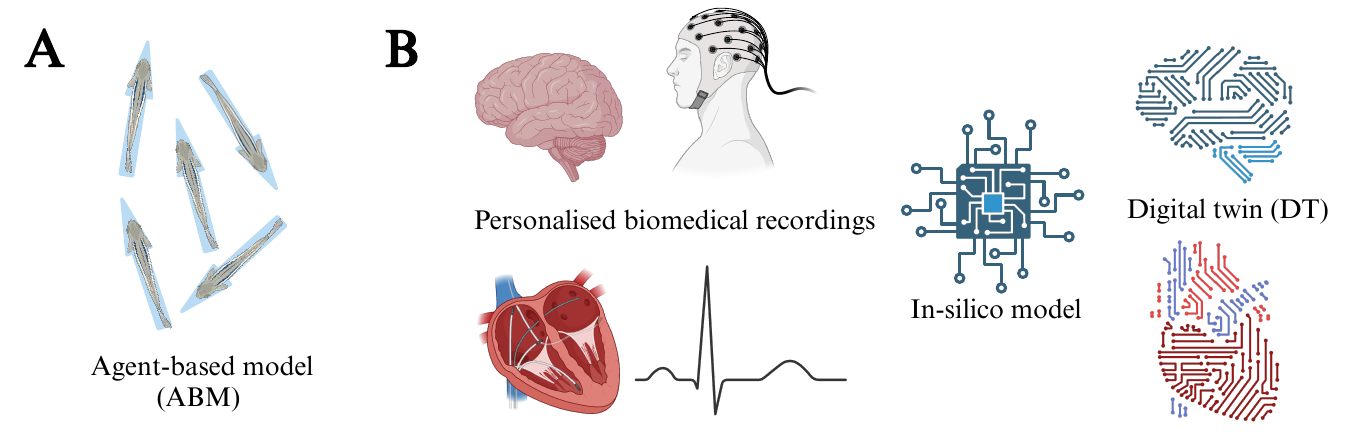}
    \caption{\textbf{Simulation-based modelling.} A. Agent-based models are powerful tools for describing the behaviour of heterogeneous populations of agentic units. Examples include the collective dynamics of swarms and flocks, where each agent can be modelled with its own dynamical properties. B. Digital twins are simulation-based models of physical systems implemented in a computer. One application of biomedicine is to use personalised recordings of neural activity or heartbeat data to calibrate individual in-silico models. Ultimately, this will lead to a virtual version of the biological system where interventions can be tested via simulation.}
    \label{fig: sim}
\end{figure*}
Another simulation-based framework for the modelling of complex systems, is the \textit{digital twin} (DT). A DT is an adaptive model that is designed to  virtually emulate the behaviour of a physical system \cite{Gelernter1991mirrorworlds}. Typically such a model is implemented on a computer. DTs have been considerably useful in engineering where they can be used to simulate the behaviour of a system under a range of conditions. Successful examples include spacecraft during the Apollo missions \cite{Allen2021nasa}, as well as other mechanical and aerospace engineering systems \cite{Ferrari2024digital}. In some cases, a DT is composed of an ABM, such as in models of cities. Similarly to ABMs, whilst DTs have an underlying mathematical formulation, they require exhaustive simulation and HPC. Moreover, they are often characterised by more engineered interfaces such as \textit{virtual} or \textit{augmented reality} (VR/AR), for use by stakeholders other than the mathematical modeller.\\
\\
In biology, DTs have sparked particular interest in the field of \textit{precision medicine} \cite{Collins2015precision}. Precision medicine refers to a treatment paradigm that considers the individual characteristics and data of a patient. This can be realised through the use of a personalised DT that uses biomedical data to calibrate a simulation of a specific patient's system to test interventions \textit{in-silico}\footnote{As opposed to experiments \textit{in-vitro} (in glass i.e. in a laboratory) or \textit{in-vivo} (in a living system), \textit{in-silico} refers to a third paradigm where experiments can be run on a computer.}, as illustrated in Panel B, Fig. \ref{fig: sim}. Two particular systems showing great promise are the heart \cite{Coorey2022hearttwin} and the brain \cite{Jirsa2023epilepsy}. In particular, \textit{The Virtual Brain} (TVB), more specifically \textit{The Virtual Epileptic Patient} (VEP), is a project leveraging detailed biophysical models and HPCs, that predicts personalised estimates of epileptogenic zone networks \cite{Wang2023epilepsy}. The empirical, actionable success of this approach stands in contrast with the dominant paradigm of complex systems neuroscience that focuses on simple oscillatory models of neural dynamics \cite{Ashwin2016oscillatory} and has yet to realise a concrete medical application \cite{Turkheimer2022complexsystems}. Nevertheless, it would be naive to assume that the pure emulation of a system is either sufficient or necessary to understand it. \textit{The Human Brain Project}\footnote{Continuing now as the \textit{Blue Brain Project}.}, was one such attempt to emulate the whole brain from individual neurons upwards \cite{Markram2006bluebrain}, that, despite some notable results at the micro-scale \cite{Markram2015reconstruction}, completely failed to deliver on its original promises \cite{Theil2015humanbrainproject}. \Rev{On the other hand, whilst models like the VEP can have strong predictive power, this may come at the cost of interpretability. Important neuroscientific phenomena such as synchronisation or state-switching can be obscured by the more complex model whilst they would be obvious in an oscillator network or hidden Markov model. Moreover, they require large datasets of personalised neural recordings, therefore relying primarily on EEG, as opposed to more expensive, but much richer, modalities such as fMRI or MEG\footnote{Electroencephalography (EEG), functional magnetic resonance imaging (fMRI) and magnetoencephalography (MEG) are methods of recording neural activity in humans.}. Finally, the VEP requires large computational resources, as opposed to simplified models, which can be yield some interesting and important insights at the `laptop-level'.}\\
\\
At their best, DTs incorporate the necessary complexity to answer specific questions, test interventions, and simulate different environments, limited by the available data, rather than a perfectly faithful reconstruction of an entire system which achieves very little in scientific terms.\\
\\
Both DTs and ABMs represent a new perspective to modelling that relies on computational simulation and empirical data. Blending this approach with modern technology, such as VR and AR, can lead to novel discoveries, out of the reach of simple mathematical models. To illustrate this, we highlight a recent study by Sayin et al \cite{Sayin2025locusts} that investigated collective behaviour by analysing real locusts interacting in VR with a swarm of simulated counterparts. This experiment highlighted the insufficiency of the elegant, dominant, but reductionist, \textit{Vicsek flocking model} \cite{Vicsek1995selfdriven} and brought new insight into the collective motion of swarming insects. At its core, this approach remains MM. However, the main-streaming of such frameworks requires the uprooting of the long-entrenched dogmas in applied mathematics that favour elegant solutions over realism and complexity.

\subsection{Automatic modelling}
\Rev{Advances in simulation-based science open the door for a new paradigm that we call} \textit{automatic modelling} (AM). \Rev{By combining techniques in ML, system identification, and model selection}, \Rev{AM is an encompassing term for methods that are able to leverage simulations to perform `theory-discovery' alongside \textit{a posteriori reductionism}. We envisage a framework where modellers initially construct detailed models that incorporate all available complexities. Subsequently, employing optimisation strategies from computational statistics and ML, these models are systematically simplified by fitting them to experimental data while simultaneously penalising model complexity. This approach would allow AM to extract, in a sparse manner, the most relevant features or underlying physical principles essential for accurately explaining the observed system behaviour. Depending on the required predictive power, users of AM could adjust model complexity against model interpretability. It would answer the crucial question: how much of the data can be explained by a reduced set of key mechanisms?}\\
\\
\Rev{Whilst this framework remains to be formally established, it shows strong alignment with existing model-selection and regularisation techniques, thus suggesting that it will become feasible within the coming decades, especially as LLM prompts for scientific discovery become increasingly conversational. Moreover, there exist some preliminary related approaches such as that of the \textit{Bayesian machine scientist} \cite{Guimerà2020bayesianmachine}, which explores and selects between candidate models using Monte-Carlo methods and Bayesian inference, penalising model-complexity, in order to perform system identification. \textit{AI-Feynman} \cite{Udrescu2020AIFeynman}, \textit{AI-Descartes} \cite{Cornelio2023aidescartes}, and \textit{AI-Hilbert} \cite{Cory2024hilbert} are three related approaches that attempt to discover physical laws directly from observed data by blending theory and experimental data using ML techniques. AM represents a future ideal where rather than relying on subjective decisions by modellers or opaque nonlinear transformations, the underlying engine could objectively identify models with essential features. This would reveal the dominant mechanism responsible for explaining experimental data through an automated process.}
\Rev{
\subsection{Generative ML and AI}
It would be remiss to discuss the future of MM without making reference to the incredible developments in generative forms of ML and AI. In particular, innovations such as the transformer architecture \cite{Vaswani2017attention}, derivatives such as large language models (LLM), or other techniques, such as diffusion models \cite{Sohl2015diffusion}, seem posed to revolutionise society, including science, even after taking into account the unreasonable media hype. In particular, generative ML stands in stark contrast with the discriminative models of the past, as these methods learn a dynamical process that is capable of sampling new candidates, including protein structures \cite{Jumper2021alphafold}, molecules \cite{Martinelli2022generative}, and human-level language capabilities \cite{GPT4tech}.\\\\
Despite their generative and predictive power, these uninterpretable, black-box models seem at odds with MM, nevertheless there are a number of complementary directions. Whilst transformers lack the biological plausibility of spiking or recurrent networks, their function resembles that of grid-cells in the hippocampus \cite{Whittington2022transformer}, suggesting that they could yet teach us something about the brain \cite{Muller2024transformers}, or simply prompt new methods in dimensionality reduction and visualisation for biological data \cite{Gilpin2025cell}. Moreover, transformers have lead to the development of so called \textit{foundation models}, which can subsequently be fine-tuned for a number of scientific tasks, as diverse as predicting chaotic \cite{Lai2025panda} and stochastic dynamics \cite{casert2024learning} or analysing single-cell data \cite{Chevaliera2025teddy}. As such algorithms develop, they become a new tool in the arsenal of the modern mathematical modeller, opening up a plethora of synergistic methods \cite{Noordijk2024rise}, including improved model calibration or data-driven approaches to dynamical systems.
}

\section{Modelling as an inverse-problem}
The traditional approach to modelling empirical observations of a dynamical system is to attempt to write down differential equations that describe the behaviour of the system. Next, one attempts to find the parameters of this model that best explain the available data. However, in large-scale complex systems, writing down a mechanistic model becomes a difficult challenge. For example, whilst Hodgkin and Huxley were able to isolate and the study the dynamics of a single neuron in a model organism, the same cannot be performed for a macroscopic brain region in a human. However, modellers now have access to an unprecedented amount of empirical data i.e. neural recordings. This has inspired the development of a modern \textit{data-driven} approach to dynamical systems, specifically the area of \textit{system identification} \cite{brunton2022datadriven}. In particular, both parameter inference and system identification are  examples of an \textit{inverse problem} \cite{Stuart2020inverse}, which involves calculating a set of causal factors, be it parameter- or model-selection, from a set of observations.\\
\\
These techniques usually begin with a \textit{multivariate time-series} (MVTS), $\{x_i(t_j):i=1,..,N, t_j = j\Delta t \}$, which records snapshots of the state, $\mathbf{x}=(x_1,..,x_N)$, at discrete time-points, $t_j$. Typically, we assume that the time-series was generated from a dynamical system of the form,
\begin{align}
    \frac{d\mathbf{x}}{dt}& = \mathbf{f}(\mathbf{x},t) + \xi(\mathbf{x},t),
\end{align}
where $\mathbf{f}$ represents the mechanistic dynamics of the system, known as the \textit{drift}, and $\xi$ represents the additive noise, known as the \textit{diffusion}. The goal of a data-driven method is to infer $\mathbf{f}$ from the MVTS\footnote{Stochastic methods often also attempt to infer the diffusion tensor $\mathbf{D}(\mathbf{x},t) = \frac{1}{2}\xi(\mathbf{x},t)\xi^{\top}(\mathbf{x},t)$ \cite{Friedrich2011complexity}.} \cite{Friedrich2011complexity}. Here methods can vary, with some attempting to infer the exact mathematical terms in $\mathbf{f}$, typically known as \textit{sparse regression} methods, and others approximating $\mathbf{f}$ with deep learning methods. Popular exisiting techniques include \textit{SINDy} \cite{Brunton2016sindy}, which performs sparse regression for nonlinear, deterministic dynamics, as well as its extensions to stochastic differential equations (SDEs) \cite{Callaham2021stochasticsindy} and PDE \cite{Rudy2017sindypde} models. For SDEs, a number of other methods exist such as \textit{stochastic force inference} \cite{Frishman2020stochasticforce,bruckner2020inferring}, \textit{InferenceMAP} \cite{elbeheiry2015inferencemap} and maximum-likelihood estimation \cite{Friedrich2011complexity}. These methods have particular applications to Brownian dynamics in soft-matter systems \cite{garcia2018reconstruction,bruckner2024learning,Gnesotto2020learning,Ronceray2024learning}. More recently, extensions have been considered for high-dimensional networked systems \cite{Gao2024learning,Gao2022autonomous}.\\\\
Beyond sparse force inference methods, the dynamics, $\mathbf{f}$, can be approximated using \textit{generative machine learning} \cite{Gilpin2024generative}. Deep learning architectures such as \textit{recurrent neural networks}\footnote{\textit{Reservoir computers \cite{Gauthier2021nextgen}} are also often used for predicting nonlinear dynamics, however they are a form of RNN.} (RNNs) and \textit{transformers} have been used to learn and predict dynamics from observations with applications to neural and molecular data \cite{Durstewitz2023reconstructing,casert2024learning,Koppe2019identifying}.
\subsection{Latent representations for dynamical systems}
\begin{figure*}
    \centering
    \includegraphics[width=\linewidth]{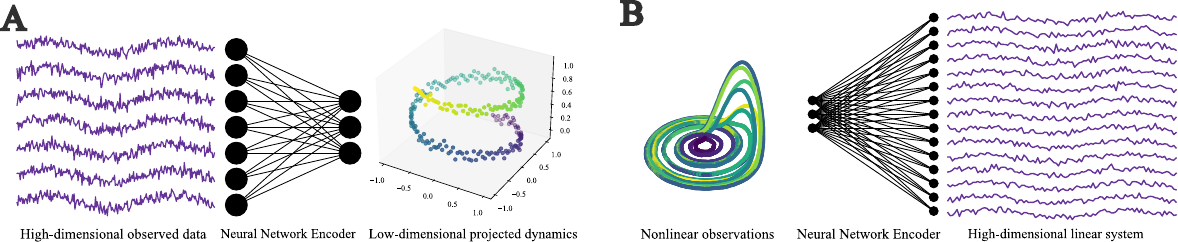}
    \caption{\textbf{Latent representations for observed dynamical systems.} A. High-dimensional data from complex biological systems is often clustered onto a low-dimensional manifold. Using data science techniques such as variational auto-encoders, we can discover the low-dimensional latent dynamics of the process. B. Following the theory of the Koopman operator, nonlinear dynamics can be `lifted' into high-dimensional spaces where the dynamics are approximately linear. This projection can be achieved with deep learning techniques such as neural networks.}
    \label{fig: latent}
\end{figure*}
Many methods for analysing nonlinear dynamics from high-dimensional data rely on the \textit{manifold hypothesis} \cite{fefferman2016manifold}, which  argues that high-dimensional data arising from complex systems are typically clustered around a lower-dimensional manifold \cite{thibeault2024lowrank,prasse2022predicting}. As a result, the dynamics evolve in a latent space with fewer dimensions. This hypothesis justifies the use of dimensionality reduction methods \cite{vandermaaten2009dimensionality} such as \textit{Principal Component Analysis} (PCA), $t$-\textit{distributed Stochastic Neighbour Embedding} (t-SNE), and \textit{Uniform Manifold Approximation and Projection} (UMAP), which are ubiquitous in biological data analysis. Such an assumption is also prevalent in data-driven approaches to dynamical systems, where much focus is placed on learning low-dimensional latent representations of high-dimensional dynamics. Architectures such as \textit{variational auto-encoders} (VAEs) are particularly adept at projecting high-dimensional time-series into low-dimensional spaces by learning nonlinear transformations \cite{Girin2021dynamicalvae}, as illustrated in Panel A, Fig. \ref{fig: latent}. Examples of this approach include the discovery of whole-brain dynamics from neural recordings, which has been used to develop heterogenous models of brain regions \cite{Sip2023characterising}.\\\\
Whilst latent dimensions no longer correspond to physical measurements from a system, they can remain interpretable. For example, in the study of motor-control, so called \textit{neural manifolds} are hypothesised to encode representations of physical space in neural activity \cite{Gallego2017manifolds}, which can be decoded using deep learning \cite{Gosztolai2025interpretable}. Latent variables can also be constrained to discrete-state spaces such as \textit{hidden Markov models} (HMMs). HMMs can be fit to time-series data using variational inference \cite{Beal2003variational} or \textit{VAMPNets} \cite{mardt2018vampnet}. These have been applied extensively in biology to uncover discrete latent states, such as in genomic sequence analysis \cite{Yoon2009hmms}, or in neural recordings \cite{viduarre2017hierarchical,Vidaurre2018spontaneous}.\\\\
Opposite to dimensionality reduction, \textit{`lifting'}, projection of dynamics into higher-dimensional space, is also particular useful for solving inverse problems. In particular, the \textit{Koopman operator} lifts nonlinear dynamics into an infinite-dimensional system where the dynamics are linear \cite{brunton2022koopman,Mezíc2021Koopman,Mezíc2024Operator}. Finite approximations of the Koopman operator can be learned using \textit{dynamic mode decomposition} \cite{Williams2015koopman} and deep learning \cite{Lusch2018linear}, as illustrated in Panel B, Fig. \ref{fig: latent}.\\\\
Data-driven approaches to dynamical systems represent a modern framework for the modelling of complex systems. When dealing with high-dimensional and complex systems, the construction of mechanistic dynamical models from first principles has proved an insurmountable challenge. On the other hand, data-driven methods leverage the enormous amounts of available data, alongside modern ML techniques, to reframe modelling as a data-led, inverse problem. Such techniques are not only predictive, but often informative and interpretable. For example, lifted linear embeddings of nonlinear dynamics are more tractable for mathematical analysis, such as eigendecomposition, than their nonlinear counterparts. Moreover, low-dimensional latent dynamics can give insight into the effective dynamics of complex systems that are obscured by dimensionality. The modelling of complex biological systems can benefit greatly from an inferential approach that focuses on predicting observed dynamics rather than theoretical models, particularly towards the aim of system control through perturbation and intervention.
\section{A fundamental physics for biology}
We have argued that theoretical biology should turn towards complex, empirical models with predictive capabilities, even if these models occasionally complicate traditional mathematical analyses. Nevertheless, this shift does not negate the importance of identifying fundamental theoretical principles that govern biological systems\Rev{, but instead, supports it}. Unlike physics, biology still lacks clear translational, foundational principles. Thus, the field urgently requires the development of a 'theoretical physics for living systems' \cite{Mehta2024theory}. Whilst significant progress remains limited, several promising directions have emerged in recent years\Rev{, where holistic MM will play a crucial role}.\\
\\
Models of collective phenomena often exhibit \textit{phase-transitions} occurring at \textit{critical points} \cite{Stanley1987phasetransitions}. In 1987, Bak et al. hypothesised that complex systems may \textit{self-organise} to the critical point, where both correlation-length across the system and sensitivity to perturbations are maximised \cite{Bak1987selforganised,Bak1995complexity}. This theory sparked particular interest in biological sciences, where evidence of \textit{self-organised criticality} has been found in neuronal firing \cite{Beggs2003avalanches} and genetic expression \cite{Vidiella2021engineering,Nykter2008genecriticality}, leading to more general theories about how critical dynamics facilitate biological function \cite{Muñoz2018criticality,Mora2011biologicalcriticality}. Moreover, recent work formalising the concept of \textit{Griffith's phases}, parameter regions showing critical-like dynamics as opposed to isolated points, promise new avenues for the furtherance of this theory \cite{Moretti2013griffiths}, which could prove to be both a governing principle for biological systems, and explain observed biological complexity. \Rev{The study of criticality in neural systems has motivated a number of data-driven approaches to dynamical systems, in particular to the problem of inferring collective properties from a fraction of total observables \cite{Levina2022subsampling,Wilting2018inferring} - techniques with implications far beyond neuroscience.}\\
\\
Another prominent theory accounting for self-organisation in biological systems is the \textit{free-energy principle} (FEP) \cite{Friston2010fep}. Originally proposed as a theory of computation in the brain, the FEP suggests that biological organisms encode sensory information in the form of a probabilistic world model that is updated according to Bayesian mechanics\footnote{It is worth noting that the `energy' in the FEP is not usually a physical energy but a larger unifying quantity that plays a similar role.}. Next, it posits that organisms act in order to minimise the \textit{surprise}\footnote{The negative log probability of an outcome.} i.e. the difference between sensory information and the world model. Whilst the FEP is formulated to account for perception-actions loops in the brain, it represents a more general theory of decision making that has been recently used to model collective behaviour in swarms \cite{Heins2024collective}. More generally, this theory, amongst other theories of agentic decision making, can be used to construct complex interacting models\Rev{, such as ABMs and DT}, that begin from a cognitive, mechanistic basis.\\
\\
Nonequilibrium thermodynamics has long been regarded as central to an understanding of the energetic constraints of biological processes \cite{Schrodinger1944whatislife,Nicolis1977self}. However, the theoretical formalism for describing the \textit{stochastic} thermodynamics of fluctuating, microscopic processes is particularly recent \cite{seifert2012thermodynamics,Broedersz202225years}.
Subsequently, stochastic thermodynamics has become extremely useful for describing the activity of biological processes, inspiring a plethora of techniques for analysing the thermodynamics of observed biological activity \cite{battle2016brokendetailedbalance,Li2019quantifying,diterlizzi2024variancesum,Horowitz2019uncertainty}, with recent interest in neural activity \cite{lynn2021detailedbalance,nartallokalu2025multilevel,nartallokalu2025review}. \Rev{In particular, a major focus has been placed on the reconstruction of force fields from stochastic trajectories, to estimate thermodynamic quantities \cite{Frishman2020stochasticforce,nartallokaluarachchi2024decomposing}. In addition, recent work has shown that ML algorithms can learn thermodynamic laws, such as the arrow of time, opening new avenues for the analysis of biological data \cite{Seif2021machinelearning}}. Despite these advancements, there remain a number of open questions regarding the thermodynamic limits of complexity and information processing \cite{Wolpert2019thermodynamics}, and how this pertains to biological processes. Nevertheless, investigating the thermodynamic constraints on biological processes promises to be an avenue ripe for discovery.\\
\\
Finally, in an effort to describe the mechanisms by which complex structures, such as amino acids, evolve, Sharma et al. recently proposed \textit{assembly theory} \cite{Sharma2023assembly}. Assembly theory quantifies the complexity of a structure in terms of the amount of selection required to produce it. Whilst theoretical in nature, assembly theory offers an avenue for understanding the constraints on the complexity of molecules, giving insight into evolution, but also yielding a testable indicator for the existence of extraterrestrial life \cite{Marshall2021assembly}. \Rev{Moreover, its algorithmic nature makes it well suited for simulation-based discovery and in-silico experimentation, ultimately offering new directions towards artificial or digital life \cite{Walker2013algorithmic}}.\\
\\
In the near future, mathematical biology and complex systems should prioritise developing empirically predictive and practical models, while continuing the pursuit of underlying theoretical principles. This area of investigation requires bold theorising that, whilst controversial at times, promises to move the field forward. Modellers should err towards theoretical predictions that are easily testable on existing data, and are rooted in established and well-tested physical principles. Progress on this front will only serve to improve the mechanistic understanding and empirical usefulness of the complex models that we advocate for. Again, these goals may be symbiotic. As mathematical and computational models begin to yield useful, actionable insights, the apparent resistance in empirical biological sciences to theoretical hypotheses may shift, ultimately blurring the edges between domains and leading to a truly interdisciplinary approach. \Rev{Moreover, the holistic MM advanced here will naturally lead to the development of further theories. For example, system identification will reveal novel mechanisms governing biological processes, whilst simulation-based approaches amount to in-silico experimentation where theories can be efficiently tested and remoulded. Together, these techniques yield a novel approach to science that is better fit for both our unsolved problems and the modern world.}

\section{Back to the future}

While it is foolish to predict the future, it is childish to deny its inevitability. As we observe the fast developments of complexity science, perspicuous challenges emerge. Clearly, developing a holistic framework for modelling complex biological systems that explains empirical data and yields useful predictions, remains one of the most daunting yet pressing challenges in interdisciplinary science. In this Perspective, we lay out a roadmap of promising research directions and necessary conceptual shifts required to achieve significant advances in the field over the coming decades. The unique challenges of biological complexity call for innovative tools that move beyond the prevalent reductionist approaches in current mathematical modelling. We highlight three particularly promising avenues that have the potential to significantly enhance mathematical biology, by adopting a holistic viewpoint integrating multiple biophysical processes and scales within a single model.\\
\\
First, we advocate adopting rich and sophisticated modelling frameworks, such as multilayer, annotated, temporal, and hyper-networks, which possess sufficient expressive power to represent biological complexity accurately. Second, we recommend the adoption of simulation-based methodologies, including agent-based models and digital twins, to incorporate detailed mechanisms and rigorously test interventions. Third, we encourage the use of data-driven modelling for dynamical processes, emphasising the inverse problem—inferring underlying mechanisms from empirical data—over the traditional forward-problem approach typically used in mathematical modelling. \\
\\
Nonetheless, we reaffirm that the identification of fundamental theoretical principles should remain the ultimate goal of biology. We highlight recent key advances offering translational, foundational insights into longstanding biological questions, building upon established theories from physics. \Rev{We believe that holistic mathematical modelling promises to yield new paradigms capable of constructing and testing biological theories efficiently, hence achieving the dream of a truly quantitative theory of life.}\\
\\
One can appreciate the substantial progress achieved over recent decades while recognising that significant further advances are both attainable and essential. Successfully addressing these challenges would dramatically transform interdisciplinary research in mathematics and biology, paving the way for breakthroughs in synthetic biology, artificial life, and numerous other biotechnologies currently confined to science fiction.
\section*{Author contributions}
R.N.-K, R.L and A.G designed the study. R.N.K wrote the manuscript. R.N.-K, R.L and A.G reviewed and edited the manuscript.
\section*{Acknowledgements}
R.N.-K. is supported by an Engineering and Physical Sciences Research Council (EPSRC) doctoral scholarship from grants EP/T517811/1 and EP/R513295/1, and The Alan Turing Institute through an Enrichment Community Award which is funded under grant EP/Z532861/1. R.L. is supported by the EPSRC grants EP/V013068/1, EP/V03474X/1 and EP/Y028872/1.

\bibliographystyle{IEEEtran} 
\bibliography{Bibliography}

\end{document}